\documentclass[12pt,thmsa]{article}
\usepackage{sw20lart}

\input{tcilatex}
\begin{document}

\title{Preliminary analysis of a recent experiment by F. A. Bovino}
\author{Emilio Santos \\
Departamento de F\'{i}sica. Universidad de Cantabria. \\
Santander. Spain}
\maketitle

\begin{abstract}
An analysis is made of the results of a recent polarization correlation
experiment by Bovino (unpublished) where about 60,000 data have been
obtained. I assume that the state of the photon pairs produced in the source
(a non-linear crystal) are in a (sightly) non-maximally entangled state and
the most relevant non-idealities of the set-up are taken into account. A
comparison is made between the predictions of quantum mechanics and a simple
family of local hidden variables models with the result that the former is
violated by more than 4 standard deviations whilst the data are compatible
with the family of local models.
\end{abstract}

The possibility of making a true discrimination between local realism and
quantum mechanics, via loophole-free tests of Bell\'{}s inequality, seems
extremely difficult or impossible as shown by the fa\"{i}lure of the efforts
made during more than 40 years after Bell\'{}s work. Thus the standard
practice has been to test inequalities derived from local realism \textit{%
plus} the assumption of \textit{fair sampling}. These inequalities have been
violated in many experiments\cite{Genovese}. However fair sampling is an
assumption theoretically absurd (because it excludes all sensible hidden
variables theories\cite{S2006}) and it has been claimed to be empirically
refuted\cite{Adenier}. Thus, in order to make real progress we should test
inequalites valid for restricted, but sensible, families of local hidden
variables (LHV) theories such as the ones derived in Ref.\cite{Santos} and
sumarized in Ref.\cite{Santos07}. In this light I will analyze the results
of a recent experiment by Bovino\cite{Bovino}.

In the Bovino experiment the detection efficiency of photon counters has
been 55\% and the overall efficiency about 17\%\cite{Bovino}. Thus, assuming
losses of order 15\% in filters\cite{Bovino2003}, I may estimate that there
is a collection efficiency (``effective fiber coupling coefficients'') about
36\%. Bovino used two-channel polarizers and he has recorded a huge amount
of data which correspond to 4 coincidence rates and 4 single rates for 45
angles of Alice\'{}s polarizers combined each with 45 angles of Bob\'{}s
polarizers.

In the ideal case quantum mechanics predicts for the coincidence rates, $%
R_{12}$, as a function of the angle, $\phi ,$ between Alice\'{}s and
Bob\'{}s polarizers, a cosinus curve of the form

\begin{equation}
R_{12}^{Q}(\phi _{j})=\left\langle R_{12}\right\rangle \left[ 1+V\cos (2\phi
)\right] ,  \label{11}
\end{equation}
The disagreement of the experimental data, $R_{12}^{\exp }(\phi _{j}),$ with
this prediction may be measured by the quantity 
\begin{equation}
\Delta =\sqrt{\sum \left[ R_{12}^{Q}(\phi _{j})-R_{12}^{\exp }(\phi
_{j})\right] },  \label{a1}
\end{equation}
where $\left\langle R_{12}\right\rangle $ and $V$ are chosen in eq.$\left( 
\ref{11}\right) $ so that $\Delta $ is a minimum. There are 4 coincidence
rates for every one of the 45 positions of Bob\'{}s polarizer, so that we
may determine $180$ quantities like $\left( \ref{a1}\right) .$ The values
obtained range between about 0.01 and 0.05 with statistical errors of order
0.01 in all cases. In about 2/3 of cases the deviation surpasses 2$\sigma $
(standard deviations) and in about 1/3 it surpasses 4$\sigma .$ So we might
conclude that there is a significant disagreement of the experimental data
with the quantum prediction, eq.$\left( \ref{11}\right) $. However it is
most appropriate to attribute the disagreement to the non-idealities of the
experimental set-up than to a true violation of quantum mechanics. Thus a
more sophisticated analysis is needed. Obviously the non-idealities should
be also taken into account in the comparison of the results with LHV models.

It is rather obvious that the experiment is not loophole-free (which would
require global detection efficiencies greater than about 80\%) so that it
does not refute the whole family of LHV theories (a family which I have
labeled LHV0 for short\cite{Santos07}). Also apparently the results do not
refute a simple family, defined in Ref.\cite{Santos}, which I have labelled
LHV1. In fact, this family predicts that the quantity $\Delta ,$ eq.$\left( 
\ref{a1}\right) ,$ should be larger than about 0.001, which is indeed the
case. In contrast the family of models labelled LHV2, which restricts LHV1
with the additional assumption of ``fair sampling'' applied to the
collection efficiency (``effective fiber coupling coefficients'') and the
filters but not to detectors, seems to be violated. It predicts that $\Delta 
$ should be larger than 0.04, a constraint not fulfilled in about half the
cases (the document Santos3.xls sent by Bovino reports that $\Delta $ -
labelled D(eta) there - is about 0.12 but my calculation gives a value about
1/3 that of Bovino). In any case the violation might be attributed to the
non-idealities of the experimental set-up.

In summary the previous (rather poor) analysis of the experiment seems to
imply that it is compatible with the family LHV1 (and therefore LHV0) but
disagrees with both the family LHV2 and quantum mechanics. In order to get
more interesting information a better analysis of the data is required,
which is made in the following.

The non-idealities of the experimental set-up are quite important, as is
shown by the fact that the 4 single rates are very different, ranging from
about 70000 to 110000 counts in the (unspecified) time window. Furthermore,
for a fixed position of Bob\'{}s analyzer Alice\'{}s single rates depend on
the angle of Alice\'{}s analyzers, with a variation up to 7\% between the
maximum and the minimum value. Similar variations exist for the single rates
when we consider different values of Bob\'{}s analyzers with Alice\'{}s
analyzers fixed. (No significant variation exists in the single counts of
Alice (Bob) when Bob (Alice) polarizer is rotated, as is expected by the
``no-signalling principle'' which forbids sending information at a
distance.) I shall make the quantum-mechanical analysis of the experiment by
studying the state produced in the non-linear crystal and how this state
evolves in the travel of photons until the detectors, as follows.

1. I should assume that in the nonlinear crystal the photons are produced
always in pairs (no single-photon productions) and that the two photons in a
pair are in some pure, entangled, quantum-mechanical state. The entanglement
may not be maximal, although close to maximal. Thus I will assume that the
state is of the form

\begin{equation}
\mid \psi >=\frac{1}{\sqrt{1+\left( 1+\gamma \right) ^{2}}}\left[ \mid
H>_{1}\otimes \mid V>_{2}-\left( 1+\gamma \right) \mid V>_{1}\otimes \mid
H>_{2}\right] ,  \label{a2}
\end{equation}
where $\gamma $ is a real number such that $\left| \gamma \right| <<1.$

2. Only a fraction $\mu _{a}$ ($\mu _{b})$ of the photons going to Alice
(Bob) are collected, so that the state of the photons in the optical fibers
becomes a statistical mixture of

a) the initial two-photon state $\left( \ref{a2}\right) $ with weight $\mu
_{a}\mu _{b},$

b) a single-photon state horizontally polarized for Alice with weight $\mu
_{a}(1-\mu _{b})/(2+2\gamma +\gamma ^{2})$ ,

c) a single-photon state vertically polarized for Alice with weight $\mu
_{a}(1-\mu _{b})(1+\gamma )^{2}/(2+2\gamma +\gamma ^{2})),$

d) a single-photon state horizontally polarized for Bob with weight $\mu
_{b}(1-\mu _{a})(1+\gamma )^{2}/(2+2\gamma +\gamma ^{2})),$

e) a single-photon state vertically polarized for Bob with weight $\mu
_{b}(1-\mu _{a})/(2+2\gamma +\gamma ^{2})$ ,

f) the vacuum state with weight $1-\mu _{a}-\mu _{b}+\mu _{a}\mu _{b}.$

3. The transmittances of the polarization analyzers of Alice for horizontal
or vertical polarization are such that, when a beam of linearly polarized
light with intensity $I_{in}$ arrives at the polarizer, the transmitted and
reflected intensities are, respectively, 
\begin{eqnarray}
I_{a+}\left( \theta \right) &=&I_{in}\left[ \left( T_{a+}-t_{a+}\right) \cos
^{2}\theta +t_{a+}\right] =\frac{1}{2}I_{in}\left[ \left(
T_{a+}+t_{a+}\right) +\left( T_{a+}-t_{a+}\right) \cos 2\theta \right] ,\ 
\label{a3} \\
I_{a-}\left( \theta \right) &=&I_{in}\left[ \left( T_{a-}-t_{a-}\right) \cos
^{2}\theta +t_{a-}\right] =\frac{1}{2}I_{in}\left[ \left(
T_{a-}+t_{a-}\right) +\left( T_{a-}-t_{a-}\right) \cos 2\theta \right] , 
\nonumber
\end{eqnarray}
and similar for Bob with the changes $a\rightarrow b.$ The sum $T+t$ is
close to unity whilst $0<t<<1$.

4. The quantum efficiencies of the 4 detectors are different, say $\zeta
_{a+}$ and $\zeta _{a-}$ for Alice and $\zeta _{b+}$ and $\zeta _{b-}$ for
Bob.

In the ideal situation the quantum prediction for the 4 coincidence
probabilities of state $\left( \ref{a2}\right) $ would be\cite{Adenier} 
\begin{eqnarray}
P_{++} &=&\frac{1}{1+\left( 1+\gamma \right) ^{2}}\left[ \left( 1+\gamma
\right) \sin \alpha \cos \beta -\cos \alpha \sin \beta \right] ^{2}, 
\nonumber \\
P_{+-} &=&\frac{1}{1+\left( 1+\gamma \right) ^{2}}\left[ \left( 1+\gamma
\right) \sin \alpha \sin \beta +\cos \alpha \cos \beta \right] ^{2}, 
\nonumber \\
P_{-+} &=&\frac{1}{1+\left( 1+\gamma \right) ^{2}}\left[ \left( 1+\gamma
\right) \cos \alpha \cos \beta +\sin \alpha \sin \beta \right] ^{2}, 
\nonumber \\
P_{--} &=&\frac{1}{1+\left( 1+\gamma \right) ^{2}}\left[ \left( 1+\gamma
\right) \cos \alpha \sin \beta -\sin \alpha \cos \beta \right] ^{2},
\label{a5}
\end{eqnarray}
where $\alpha $ ($\beta )$ is the angle of Alice\'{}s (Bob\'{}s)
polarization analyzer. Now I introduce the most relevant non-idealities as
follows. Due to losses and absorptions, as explained above, there is a
global factor 
\[
\mu _{a}(T_{a+}+t_{a+})\zeta _{a+}\mu _{b}(T_{b+}+t_{b+})\zeta _{b+} 
\]
in front of $P_{++}$ and similarly for $P_{+-},P_{-+}$ and $P_{--}$. In
addition every $\cos 2\alpha $ or $\sin 2\alpha $ should be preceded by
either a factor $(T_{a+}-t_{a+})/(T_{a+}+t_{a+})$ or a factor $%
(T_{a-}-t_{a-})/(T_{a-}+t_{a-})$ (see eqs.$\left( \ref{a3}\right) )$ and
similarly ,with $a\rightarrow b,$ for $\cos \beta $ or $\sin \beta .$ Thus I
get 
\begin{eqnarray}
P_{++} &=&\frac{1}{4}\eta _{a+}\eta _{b+}\left[ 1-V_{++}\cos 2\phi +\gamma
^{\prime }(\cos 2\beta -\cos 2\alpha )+\gamma ^{\prime \prime }\sin 2\alpha
\sin 2\beta \right] ,  \nonumber \\
P_{+-} &=&\frac{1}{4}\eta _{a+}\eta _{b-}\left[ 1+V_{+-}\cos 2\phi -\gamma
^{\prime }(\cos 2\beta +\cos 2\alpha )-\gamma ^{\prime \prime }\sin 2\alpha
\sin 2\beta \right] ,  \nonumber \\
P_{-+} &=&\frac{1}{4}\eta _{a-}\eta _{b+}\left[ 1+V_{-+}\cos 2\phi +\gamma
^{\prime }(\cos 2\beta +\cos 2\alpha )-\gamma ^{\prime \prime }\sin 2\alpha
\sin 2\beta \right] ,  \nonumber \\
P_{--} &=&\frac{1}{4}\eta _{a-}\eta _{b-}\left[ 1-V_{--}\cos 2\phi -\gamma
^{\prime }(\cos 2\beta -\cos 2\alpha )+\gamma ^{\prime \prime }\sin 2\alpha
\sin 2\beta \right] ,  \label{a6}
\end{eqnarray}
where 
\[
\phi \equiv \alpha -\beta ,\;\gamma ^{\prime }\equiv \frac{2\gamma +\gamma
^{2}}{2+2\gamma +\gamma ^{2}},\;\gamma ^{\prime \prime }\equiv \frac{\gamma
^{2}}{2+2\gamma +\gamma ^{2}},\;\eta _{a+}\equiv \mu _{a}T_{a+}\zeta _{a+}, 
\]
and similarly for the remaining parameters $\eta $. The term $V_{++}$ is
given by 
\[
V_{++}=\frac{(T_{a+}-t_{a+})(T_{b+}-t_{b+})}{(T_{a+}+t_{a+})/(T_{b+}+t_{b+})}
\]
and similarly for $V_{+-},V_{-+}$ and $V_{--}$ (see eqs.$\left( \ref{a3}%
\right) $) . The correction for finite transmittance (i. e. the fact that $%
t>0$) has not been taken into account in the terms containing the parameter $%
\gamma .$ Indeed it should be realized that $\gamma \sim 0.1$ and $t\sim
0.01,$ therefore I am neglecting terms of order $0.001$ with respect to the
main term, whilst I do not neglect terms of order $\gamma ^{2}\sim 0.01.$

Also we may calculate the probabilities for single counts by Alice, $%
P_{a+},P_{a-}$, and Bob, $P_{b+},P_{b-}$ , respectively, getting 
\begin{eqnarray}
P_{a+} &=&\frac{1}{2}\eta _{a+}\left[ 1-\gamma ^{\prime }\cos 2\alpha
\right] ,\;P_{a-}=\frac{1}{2}\eta _{a+}\left[ 1+\gamma ^{\prime }\cos
2\alpha \right] ,  \nonumber \\
P_{b+} &=&\frac{1}{2}\eta _{b+}\left[ 1+\gamma ^{\prime }\cos 2\beta \right]
,\;P_{b-}=\frac{1}{2}\eta _{b+}\left[ 1-\gamma ^{\prime }\cos 2\beta \right]
.  \label{a7}
\end{eqnarray}
The number of counts within one time window should be obtained by
multiplying the probabilities $\left( \ref{a6}\right) $ or $\left( \ref{a7}%
\right) $ times $R_{0}$, this being the number of photon pairs produced in
the source within the window. That is $R_{++}=R_{0}P_{++}$, etc.

The question is whether all the data of Bovino\'{}s experiment may be fitted
to eqs.$\left( \ref{a6}\right) $ and $\left( \ref{a7}\right) $ with 10 free
parameters, namely $\eta _{a+},$ $\eta _{b+},$ $\eta _{a-}$, $\eta _{b-}$, $%
\gamma $, $V_{++}$, $V_{+-}$,$V_{-+}$,$V_{--}$ and $R_{0}.$ The analysis
simplifies a lot as follows. From eqs.$\left( \ref{a7}\right) $ I get the
global efficiencies by averaging over angles, that is 
\begin{equation}
\eta _{a+}=2\left\langle P_{a+}\right\rangle ,\eta _{b-}=2\left\langle
P_{b-}\right\rangle ,\eta _{a-}=2\left\langle P_{a-}\right\rangle ,\eta
_{b+}=2\left\langle P_{b+}\right\rangle .  \label{c5}
\end{equation}
Now I may eliminate the efficiencies $\eta _{a+},$ etc. in eqs.$\left( \ref
{a6}\right) $ using eqs.$\left( \ref{c5}\right) $ and pass from
probabilities to count numbers by multiplication times $R_{0}$ in the
appropriate places. Thus I get 
\begin{eqnarray}
\frac{R_{0}R_{++}}{\left\langle R_{a+}\right\rangle \left\langle
R_{b+}\right\rangle } &=&1-V_{++}\cos 2\phi +\gamma ^{\prime }(\cos 2\beta
-\cos 2\alpha )+\gamma ^{\prime \prime }\sin 2\alpha \sin 2\beta ,  \nonumber
\\
\frac{R_{0}R_{+-}}{\left\langle R_{a+}\right\rangle \left\langle
R_{b-}\right\rangle } &=&1+V_{+-}\cos 2\phi -\gamma ^{\prime }(\cos 2\beta
+\cos 2\alpha )-\gamma ^{\prime \prime }\sin 2\alpha \sin 2\beta ,  \nonumber
\\
\frac{R_{0}R_{-+}}{\left\langle R_{a-}\right\rangle \left\langle
R_{b+}\right\rangle } &=&1+V_{-+}\cos 2\phi +\gamma ^{\prime }(\cos 2\beta
+\cos 2\alpha )-\gamma ^{\prime \prime }\sin 2\alpha \sin 2\beta ,  \nonumber
\\
\frac{R_{0}R_{--}}{\left\langle R_{a-}\right\rangle \left\langle
R_{b-}\right\rangle } &=&1-V_{--}\cos 2\phi -\gamma ^{\prime }(\cos 2\beta
-\cos 2\alpha )+\gamma ^{\prime \prime }\sin 2\alpha \sin 2\beta .
\label{c6}
\end{eqnarray}
Hence I may define the following average coincidence detection probability, $%
P\left( \alpha ,\beta \right) ,$ 
\begin{eqnarray}
P &\equiv &16R_{0}f  \label{d7} \\
f &\equiv &\frac{R_{+-}\left( \alpha ,\beta \right) }{\left\langle
R_{a+}\right\rangle \left\langle R_{b-}\right\rangle }+\frac{R_{-+}\left(
\alpha ,\beta \right) }{\left\langle R_{a-}\right\rangle \left\langle
R_{b+}\right\rangle }+\frac{R_{++}\left( \alpha +\frac{\pi }{4},\beta -\frac{%
\pi }{4}\right) }{\left\langle R_{a+}\right\rangle \left\langle
R_{b+}\right\rangle }+\frac{R_{--}\left( \alpha +\frac{\pi }{4},\beta -\frac{%
\pi }{4}\right) }{\left\langle R_{a-}\right\rangle \left\langle
R_{b-}\right\rangle }.  \nonumber
\end{eqnarray}
The quantum prediction for $P$ depends on the angles $\alpha $ and $\beta $
only via the combination $\alpha -\beta \equiv \phi $ and it is rather
simple, namely 
\begin{equation}
\;P^{Q}\left( \phi \right) =\frac{1}{4}\left[ 1+V\cos 2\phi \right]
,\;P\left( \phi \right) \equiv 16R_{0}f,  \label{b6}
\end{equation}
where 
\[
V\equiv \frac{1}{4}(V_{++}+V_{+-}+V_{-+}+V_{--}-\frac{2\gamma ^{2}}{%
2+2\gamma +\gamma ^{2}}). 
\]
(It may be realized that the same result is obtained putting $\left( \alpha -%
\frac{\pi }{4},\beta +\frac{\pi }{4}\right) $ as the argument of $R_{++}$
and $R_{--}$.) At this moment I stress that, for the discrimination between
quantum mechanics and LHV models, the combination of rates eq.$\left( \ref
{d7}\right) $ is more appropriate than the fashionable combination 
\begin{equation}
U(\phi )=\frac{R_{++}\left( \phi \right) +R_{--}\left( \phi \right)
-R_{+-}\left( \phi \right) -R_{-+}\left( \phi \right) }{R_{++}\left( \phi
\right) +R_{--}\left( \phi \right) +R_{+-}\left( \phi \right) +R_{-+}\left(
\phi \right) },  \label{c7}
\end{equation}
for which quantum mechanics predicts a simple expression, namely $V\cos
\left( 2\phi \right) ,$ only in the ideal case.

A good fit of the data into eqs.$\left( \ref{d7}\right) $ and $\left( \ref
{b6}\right) ,$ with $R_{0}$ and $V$ as free parameters, seems to be a
necessary condition for the compatibility of the experiment with quantum
mechanics. However, even if a good fit is not possible, still the experiment
may be compatible with quantum mechanics because the state produced in the
source may be different from eq.$\left( \ref{a2}\right) $and/or there are
additional non-idealities not included in the previous analysis.
Consequently proving an empirical violation of quantum mechanics is
extremely difficult or impossible in actual experiments. Similarly refuting
local realism is impossible whenever the global detection efficiency does
not surpasse about 80\%, a well known fact.

Nevertheless, even if the Bovino experiment\cite{Bovino} does not allow a
rigorous discrimination between quantum mechanics and local realism,
interesting information may be obtained by studying the agreement, or
disagreement, of the data with some simple LHV models departing but slightly
from the quantum predictions eqs.$\left( \ref{a6}\right) $ or $\left( \ref
{a7}\right) .$ This study would require to define a simple family of LHV
models and to find whether the data agree with either the quantum
predictions or the said simple LHV models. Constructing LHV models
appropriate for a comparison with the quantum eqs.$\left( \ref{a6}\right) $
or $\left( \ref{a7}\right) $ is a most interesting aim for the near future,
but from my experience I may guess that the results will be as follows. The
model predictions will be of the form given by eq.$\left( 37\right) $ of 
\cite{Santos}, that is (compare with eqs.$\left( \ref{b6}\right) )$

\begin{equation}
P^{LHV}\left( \phi \right) =\frac{1}{4}\left[ 1+V\cos \left( 2\phi \right)
+\delta \left( \phi \right) \right] ,  \label{b8}
\end{equation}
$\delta \left( \phi \right) $ being 
\begin{equation}
\delta \left( \phi \right) =\frac{8\varepsilon ^{3}}{3\pi }\left[ 2\frac{%
\sin ^{2}\left( \pi \eta /2\right) }{\left( \pi \eta /2\right) ^{2}}\cos
\left( 2\phi \right) -1+\frac{2}{\eta ^{2}}\left( \eta +\frac{2}{\pi }\left|
\phi \right| -1\right) _{+}\right] ,\;  \nonumber
\end{equation}
where $\phi \in \left[ -\pi /2,\pi /2\right] $ and $\left( {}\right) _{+}$
means putting $0$ if the quantity inside brackets is negative. The parameter 
$\eta $ is an averaged detection efficiency and $\varepsilon $ is the
solution of the equation 
\begin{equation}
\frac{\pi -2\varepsilon +\sin (2\varepsilon )\cos (2\varepsilon )}{\cos
(2\varepsilon )\left[ \pi -2\varepsilon +\tan (2\varepsilon )\right] }=V%
\frac{\left( \pi \eta /2\right) ^{2}}{\sin ^{2}\left( \pi \eta /2\right) }.
\label{a9}
\end{equation}
if $\varepsilon >0$ or $\varepsilon =0$ if the solution is negative.

Eqs.$\left( \ref{d7}\right) $ and $\left( \ref{b8}\right) $ may be rewritten
taking account of the first two terms of $\delta \left( \phi \right) $ by
means of small changes in the parameters $R_{0}$ and $V$, that is 
\begin{equation}
P^{LHV}\left( \phi \right) \equiv 16R_{0}^{\prime }f=\frac{1}{4}\left[
1-V^{\prime }\cos \left( 2\phi \right) +\frac{32\varepsilon ^{3}}{3\pi
^{2}\eta ^{2}}\left( \left| \phi \right| -\frac{\pi }{2}+\frac{\pi \eta }{2}%
\right) _{+}\right] .  \label{c9}
\end{equation}
After that, \textit{the discrimination between quantum and LHV predictions
will consist of checking whether the quantum eq.}$\left( \ref{b6}\right) $%
\textit{\ (with R}$_{\mathit{0}}$\textit{\ and V as free parameters) or the
LHV eq.}$\left( \ref{c9}\right) $\textit{\ (with }$R_{0}^{\prime },V^{\prime
}$\textit{\ and }$\eta $ \textit{\ as free parameters) may be fitted to the
data of the experiment.} In practice we should make chi-square fits of the
experimental data into eq.$\left( \ref{c9}\right) $ with 3 free parameters,
namely $R_{0}^{\prime },V^{\prime }$ and $\eta $. The parameter $\varepsilon 
$ is related to $\eta $ and $V^{\prime }$ by eq.$\left( \ref{a9}\right) $.
If $\eta <<1$ eq.$\left( \ref{a9}\right) $ may be solved to order $%
\varepsilon ^{2}$ giving\cite{Santos} 
\begin{equation}
\varepsilon \simeq \sqrt{\frac{1}{2}\left( 1-\frac{\sin ^{2}(\pi \eta /2)^{2}%
}{V^{\prime }(\pi \eta /2)^{2}}\right) _{+}}\simeq \sqrt{\frac{1}{2}\left(
V^{\prime }-\frac{\sin ^{2}(\pi \eta /2)^{2}}{(\pi \eta /2)^{2}}\right) _{+}}%
,  \label{a10}
\end{equation}
where the second equality is valid for $V^{\prime }$ close to unity, as is
usually the case. I guess that fairly good fits exist for some set of values
of $\eta $. If a good fit is possible for $\eta =0$ (in this case the LHV eq.%
$\left( \ref{c9}\right) $ become identical to the quantum eq.$\left( \ref{b6}%
\right) )$ then the experiment is compatible with standard quantum mechanics
(``standard'' means accepting the analysis leading to eqs.$\left( \ref{a6}%
\right) $ and $\left( \ref{a7}\right) ).$ If good fits are possible for $%
\eta \geq 0.17$ then the experiment is compatible with the family of local
models defined in \cite{Santos}, a family labelled LHV1 in \cite{Santos07}.
If there are good fits for $\eta \geq 0.55$ then the experiment would be
also compatible with the family defined as LHV2 in \cite{Santos07}, which is
more restrictive than LHV1. However I do not think the latter will be the
case in view of the results of the rough analysis made at the beginning of
this paper.

The fit of the data into the equations should be made for every one of the
46 sets of data corresponding to one Bob\'{}s polarizer position each,
rather than a fit of the whole set of data. The reason is that in Bell tests
it is essential that the measured rates correspond to the same production
rate in the source. I suppose that in the Bovino experiment it is much
easier to guarantee the constancy of the production rate for every Bob\'{}s
polarizer position than for the whole set of data.

In order to make a preliminary analysis of the experiment I have uses a few
data of $``fiBob"=90{{}^{o}}$. A simple consequence of the (LHV)\ eq.$\left( 
\ref{b8}\right) $ is, assuming $\eta \leq 1/2,$ 
\begin{equation}
\nu \equiv \frac{f\left( 0{}\right) +f\left( \pi /2\right) -2f\left( \pi
/4\right) }{f\left( 0{}\right) +f\left( \pi /2\right) +2f\left( \pi
/4\right) }=\frac{16\varepsilon ^{3}/(3\pi \eta )}{4+16\varepsilon
^{3}/(3\pi \eta )}\simeq \frac{4\varepsilon ^{3}}{3\pi \eta },  \label{c13}
\end{equation}
whilst standard quantum mechanics predicts $\nu =0.$ The value of $V^{\prime
}$ may be obtained from 
\[
V^{\prime }=\frac{f\left( 0{}\right) -f\left( \pi /2\right) }{f\left(
0{}\right) +f\left( \pi /2\right) }, 
\]
and I get 
\begin{equation}
V^{\prime }=0.9760,\;\nu ^{data}\approx 0.00149\pm 0.00032.  \label{c14}
\end{equation}
(In the experiment no data for the angle $\pi /4$ have been reported and I
have used averages between the angles 44${{}^{o}}$ and 46${{}^{o}}$.) In eq.$%
\left( \ref{c14}\right) $ the statistical error has been calculated from the
data for $\phi =\pi /4$, the errors in other data making a negligible
change. The result eq.$\left( \ref{c14}\right) $ suggests that\textit{\ the
experiment disagrees with standard quantum predictions} $\left( \nu
=0\right) $ by about 4.5 standard deviations. In order to study the
agreement with the family of local models LHV1 we should search for the
value of $\eta $ which gives $\nu ^{LHV}\simeq 0.015.$ We realize that this
is achieved by $\eta =0.225$, which gives $\varepsilon =$ $0.092$ using eq.$%
\left( \ref{a10}\right) ,$ whence eq.$\left( \ref{c13}\right) $ leads to
agreement sith the empirical result eq.$\left( \ref{c14}\right) .$
Consequently \textit{the data are compatible with the family LHV1} but 
\textit{refute the family LHV2 }because $\eta =0.225$ is greater than the
overall efficiency $0.17$ but smaller than the quantum efficiency of
detectors $0.55$.

As another example, a similar analysis of data of $``fiBob"=40{{}^{o}}$
gives 
\begin{equation}
V^{\prime }=0.9521,\;\nu ^{data}\approx 0.00142\pm 0.00047,  \label{c15}
\end{equation}
that is a result departing from standard quantum predictions, $\nu =0,$ by 3
standard deviations but in agreement with the family LHV1 of local models.

In summary, although a more complete analysis is needed, the data of the
Bovino experiment seem to depart from standard quantum predictions by an
amount which agrees in sign and order of magnitude with the predictions of a
simple family of LHV models\cite{Santos}. However I should not conclude that
a real violation of quantum mechanics has taken place. In fact, there may be
non-idealities of the set-up not taken into account in the
quantum-mechanical analysis leading to eqs.$\left( \ref{a6}\right) $ or $%
\left( \ref{a7}\right) $. Nevertheless if a more complete analysis confirms
that the data agree with the predictions of simple LHV models, eq.$\left( 
\ref{c9}\right) ,$ better than with the standard quantum prediction, eq.$%
\left( \ref{b6}\right) $, that would reinforce my conjecture that \textbf{%
non-idealities of experimental set-ups tend to save local realism}. If the
non-idealities may be explained within quantum mechanics, then the
conjecture would be that \textbf{quantum mechanics and local realism are
compatible at the empirical level}, Bell\'{}s theorem being true only for an
idealized (incorrect) version of quantum mechanics.


\begin{thebibliography}{9}
\bibitem{Genovese}  M. Genovese, \textit{Phys. Rep.} \textbf{413}, 319
(2005).

\bibitem{S2006}  E. Santos, \textit{Found. Phys.}\textbf{\ 34}, 1643 (2004);
c-print arXiv:quant-ph/0410193.

\bibitem{Adenier}  G. Adenier and A. Yu. Khrennikov, \textit{J. Phys. B: At.
Mol. Opt. Phys.} \textbf{40}, 131 (2007).

\bibitem{Santos}  E. Santos, \textit{Eur. Phys. Jour. D} \textbf{42}, 501
(2007) ; c-print arXiv:quant-ph/0612212.

\bibitem{Santos07}  E. Santos, \textit{Eur. Phys. Jour. D} (2007). In press.

\bibitem{Bovino}  F. A. Bovino (2007), unpublished. Private communication.

\bibitem{Bovino2003}  F. A. Bovino et al., c-print arXiv:quant-ph/0303126.
\end{thebibliography}
\end{document}